\documentclass[aps,pre,groupedaddress,showpacs,preprint]{revtex4}
\usepackage{graphicx}
\usepackage[dvips]{epsfig}
\usepackage{dcolumn}
\usepackage{subfigure}%
\usepackage{bm}
\usepackage{amssymb}
\usepackage{amsmath}

\begin{document}
\title{Shock wave formation in Rosenau's extended hydrodynamics}

\author{Carlos Escudero}

\affiliation{Departamento de F\'{\i}sica Fundamental, Universidad Nacional de Educaci\'{o}n a Distancia,
C/Senda del Rey 9, Madrid, Spain}

\begin{abstract}
We study the extended hydrodynamics proposed by Philip Rosenau [Phys. Rev. A {\bf 40}, 7193 (1989)]
in the context of a regularization of the Chapman-Enskog expansion. We are able to prove that shock waves
appear in finite time in Rosenau's extended Burgers' equation, and we discuss the physical implications
of this fact and its connection with a possible extension of hydrodynamics to the short wavelength domain.
\end{abstract}

\maketitle

The Boltzmann equation is one of the most fundamental equations in Nonequilibrium Statistical Mechanics. This equation
describes the dynamics of a rarefied gas, taking into account two basic processes: the free flight of the particles and
their collisions. Due to the difficulties that a direct treatment of this equation implies, a reduced description of
Boltzmann equation is one of the major problems in kinetic theory. Equations of hydrodynamics constitute a closed set of
equations for the three hydrodynamic fields: local density, local velocity, and local temperature. These equations
can be derived from Boltzmann equation by performing the Chapman-Enskog expansion~\cite{chapman}. This expansion is
a power series expansion in the Knudsen number, that is the ratio of the free mean path between the macroscopic length.
The first order of the expansion yields Euler equations, while the second order yields Navier-Stokes equations which
in the case of an incompressible fluid read:
\begin{eqnarray}
\nonumber
\partial_t {\bf v} + ({\bf v}\cdot \nabla){\bf v} &=&
-\nabla p + \mu \nabla^2 {\bf v} \\
\nabla \cdot {\bf v} &=& 0,
\end{eqnarray}
where $\mu$ represents the viscosity of the fluid, and is of the order of the Knudsen number. The next order in the Chapman-Enskog expansion yields the Burnett equations of hydrodynamics which are, unfortunately, invalid. To see this
more clearly consider the viscous part of the Chapman-Enskog expansion:
\begin{equation}
\label{chapman}
\epsilon ( \mu_0 \nabla^2 {\bf v} + \epsilon^2 \mu_1 \nabla^4 {\bf v} + ...),
\end{equation}
where $\mu = \epsilon \mu_0$, and $\epsilon$ is the Knudsen number.
The Burnett order implies the presence of the biharmonic term proportional to $\nabla^4$, that causes an unphysical
increase in the number of boundary conditions and rends the equilibrium unstable, among other undesirable effects. While
the Navier-Stokes equations give very accurate results in many domains, they usually fail when applied to predict the
short wavelength properties of the fluid, like, for instance, the propagation of ultrasounds within the fluid. This
makes very useful to develop a higher order description of the fluid, while the Burnett order has proven itself less
accurate than the Navier-Stokes order. This problem was partially solved by Philip Rosenau in his influencing article
of 1989~\cite{rosenau}. The idea was to regularize the Chapman-Enskog expansion using a very original comparison. First
consider the power series expansion:
\begin{equation}
\label{expansion}
\frac{1}{1-z}=1+z^2+z^4+...,
\end{equation}
where $z$ is a complex number which modulus fullfills $|z|<1$. Assuming that $\epsilon$ is small enough and taking
into account the power series~(\ref{expansion}) suggest that we can recast expansion~(\ref{chapman}) into the form
\begin{equation}
\frac{\mu \nabla^2}{1 - \epsilon^2 m^2 \nabla^2} {\bf v},
\end{equation}
where $m^2 = \frac{\mu_1}{\mu_0}$, and this operator is to be interpreted in the Fourier transform sense:
\begin{equation}
\left( \frac{\mu \nabla^2}{1 - \epsilon^2 m^2 \nabla^2} {\bf v} \right)^{\hat{}} =
\frac{- \mu k^2}{1 + \epsilon^2 m^2 k^2} \hat{{\bf v}}.
\end{equation}
This idea was originally proposed in the context of random walk theory~\cite{doering}, and has been used within this
context in latter works~\cite{kevrekidis}.

While this regularization of the Chapman-Enskog expansion seems to be a proper extension of hydrodynamics in the linear
regime~\cite{rosenau}, its effect on the full nonlinear hydrodynamics is not so clear. This is due to the analytical
difficulties that a mathematical treatment of the Navier-Stokes equations implie. However, it is useful to study
some toy models to win a deeper understanding of hydrodynamics; to this end was developed a one-dimensional model for
hydrodynamics: the Burgers' equation
\begin{equation}
\partial_t u + u \partial_x u = \mu \partial_x^2 u.
\end{equation}
In the same spirit, Rosenau considered the regularized Burgers' equation, arguing that an understanding of this model would
clarify the effect of the regularization of the Chapman-Enskog expansion on the nonlinear hydrodynamics. The rest of this
work is devoted to prove the appearence of shock waves in finite time in Rosenau's regularized Burgers' equation, and
to analize the physical implications of this fact.

Rosenau's extended Burgers' equation reads:
\begin{equation}
\label{eqrosenau}
\partial_t u + u \partial_x u = \mu \frac{\partial_x^2}{1-\epsilon^2\partial_x^2} u,
\end{equation}
where we have set, without lost of generality, $\mu_1/\mu_0=1$.
To prove shock wave formation we will exploit the analogy between viscous Burgers' equation
(the inviscid Burgers' equation is obtained just by setting $\mu=0$)
and the Keller-Segel
system~\cite{keller}:
\begin{eqnarray}
\label{keller1}
\partial_t v &=& \mu \partial_x^2 v + \partial_x ( v \partial_x w), \\
\partial_x^2 w &=& -v.
\label{keller2}
\end{eqnarray}
Note that we recover viscous Burgers' equation performing the substitution $u = \partial_x w$ in
system~(\ref{keller1},\ref{keller2}). Consider now the following modified Keller-Segel system:
\begin{eqnarray}
\label{kelros1}
\partial_t v &=& \mu \frac{\partial_x^2}{1-\epsilon^2 \partial_x^2} v + \partial_x ( v \partial_x w), \\
\partial_x^2 w &=& -v.
\label{kelros2}
\end{eqnarray}
We can recover Rosenau's extended Burgers' equation by performing again the substitution $u = \partial_x w$ in this
last system. We will consider homogeneous Dirichlet boundary conditions:
$v|_{\partial \Omega}=w|_{\partial \Omega}=0$,
where $\Omega$ is the closed interval $\Omega=[-L,L]$. From system~(\ref{kelros1},\ref{kelros2}) we get:
\begin{eqnarray}
\nonumber
\frac{1}{2} \frac{d}{dt} ||v(\cdot,t)||_{L^2(\Omega)}^2= \int_\Omega vv_t dx = \mu \int_\Omega
v\frac{\partial_x^2}{1-\epsilon^2\partial_x^2}v dx \\
- \int_\Omega v \partial_x w \partial_x v dx + \int_\Omega v^3 dx
\label{estimate}
\end{eqnarray}
Now, we are going to estimate all the terms appearing in the right hand side of this equation.

Integrating by parts the second term in the right hand side of Eq.(\ref{estimate}):
\begin{eqnarray}
\nonumber
\int_\Omega v \partial_x w \partial_x v dx = \left. v^2 \partial_x w \right|_{\partial \Omega}
- \int_\Omega \partial_x v \partial_x w v dx \\
- \int_\Omega v \partial_x^2 w v dx,
\end{eqnarray}
that implies
\begin{eqnarray}
\nonumber
\int_\Omega v \partial_x w \partial_x v dx = -\frac{1}{2} \int_\Omega v^2 \partial_x^2 w dx = \\
\frac{1}{2} \int_\Omega v^3 dx
\end{eqnarray}
The first term in the right hand side of Eq.(\ref{estimate}) can be estimated as follows:
\begin{eqnarray}
\nonumber
\int_\Omega v \frac{\partial_x^2}{1-\epsilon^2 \partial_x^2} v dx \le
\left| \int_\Omega v \frac{\partial_x^2}{1-\epsilon^2\partial_x^2} v dx \right| \le \\
\int_\Omega \left| v \frac{\partial_x^2}{1-\epsilon^2\partial_x^2} v \right| dx \le
\left| \left| v \right| \right|_{L^2(\Omega)} \left| \left| \frac{\partial_x^2}
{1-\epsilon^2\partial_x^2} v
\right| \right|_{L^2(\Omega)},
\end{eqnarray}
where we have used H\"{o}lder's inequality (see below). By performing the shift of variables $y=x/\epsilon$, we get:
\begin{eqnarray}
\nonumber
\left| \left| \frac{\partial_x^2}{1-\epsilon^2 \partial_x^2} v \right| \right|_{L^2(\Omega)} =
\frac{1}{\epsilon^{(3/2)}} \left| \left| \frac{\partial_y^2}{1-\partial_y^2} v \right|
\right|_{L^2(\Omega/\epsilon)} \le \\
\frac{N}{\epsilon^{(3/2)}} \left| \left| v \right|
\right|_{L^2(\Omega/\epsilon)},
\end{eqnarray}
where $N=\left| \partial_y^2(1-\partial_y^2)^{-1} \right|$.
Let us clarify a bit this last step.
We have used the fact that the operator $\nabla^2(1-\nabla^2)^{-1}$ is bounded
on every $L^p$ space, with $1 \le p \le \infty$. This means that we can assure that
$\left| \left| \nabla^2(1-\nabla^2)^{-1} f \right| \right|_{L^p(\Omega)} \le N \left| \left| f \right| \right|_{L^p(\Omega)}$
for every $f$ belonging to $L^p(\Omega)$ and a constant $N$ that does not depend on $f$ (and thus $N$ is called the norm of the operator). This fact can be easily seen
once one realizes that the Fourier transform of the operator $\nabla^2(1-\nabla^2)^{-1}$ is a bounded function of the wavevector, and a rigorous proof can be found in
~\cite{stein}.
We can again shift variables $x=\epsilon y$
to get:
\begin{equation}
\int_\Omega v \frac{\partial_x^2}{1-\epsilon^2 \partial_x^2} v dx \le
\frac{N}{\epsilon^2} \left| \left| v \right| \right|_{L^2(\Omega)}^2.
\end{equation}
Finally, we can conclude our estimate as follows:
\begin{eqnarray}
\nonumber
\int_\Omega v \frac{\partial_x^2}{1-\epsilon^2 \partial_x^2} v dx \ge
-\left|\int_\Omega v \frac{\partial_x^2}{1-\epsilon^2 \partial_x^2} v dx \right| \ge \\
- \frac{N}{\epsilon^2} \left| \left| v \right| \right|_{L^2(\Omega)}^2.
\end{eqnarray}
Now we are going to estimate the third term in Eq.(\ref{estimate}):
\begin{equation}
\int_\Omega v^3 dx = \left| \left| v \right| \right|_{L^3(\Omega)}^3.
\end{equation}
H\"{o}lder's inequality reads (for a rigorous proof of H\"{o}lder's inequality see~\cite{evans}):
\begin{eqnarray}
\nonumber
\int_\Omega \left| f g \right| dx \le \left| \left| f \right| \right|_{L^p(\Omega)} \left| \left| g \right|
\right|_{L^q(\Omega)}, \\
\qquad 1 \le p,q \le \infty, \qquad \frac{1}{p}+\frac{1}{q}=1.
\end{eqnarray}
Choosing $g=1$ we get:
\begin{equation}
\int_\Omega \left| f \right| dx \le C \left| \left| f \right| \right|_{L^p(\Omega)},
\end{equation}
where $C=\left| \Omega \right|^{1/q}$. With this estimate we can claim that:
\begin{eqnarray}
\nonumber
\left| \left| v \right| \right|_{L^2(\Omega)}^2 = \int_\Omega v^2 dx \le C \left| \left| v^2
\right| \right|_{L^p(\Omega)} = \\
\nonumber
C \left( \int_\Omega v^{2p} dx \right)^{(1/p)} = \\
C \left( \int_\Omega v^{3} dx \right)^{(2/3)} = C \left| \left| v \right| \right|_{L^3(\Omega)}^2,
\end{eqnarray}
where we have chosen $p=3/2$ (and correspondingly $q=3$). This implies that:
\begin{equation}
\left| \left| v \right| \right|_{L^3(\Omega)} \ge D \left| \left| v \right| \right|_{L^2(\Omega)},
\end{equation}
where $D=\left| \Omega \right|^{-1/6}$.
Therefore, we have the final estimate:
\begin{equation}
\frac{d}{dt} \left| \left| v \right| \right|_{L^2(\Omega)}^2 \ge
A \left( \left| \left| v \right| \right|_{L^2(\Omega)}^2 \right)^{(3/2)}-B
\left| \left| v \right| \right|_{L^2(\Omega)}^2,
\end{equation}
where $A,B>0$ are constants, $A=|\Omega|^{-1/2}$ and $B=\frac{2N\mu}{\epsilon^2}$.
We are thus going to study the dynamical system:
\begin{equation}
\frac{dx}{dt}=Ax^{3/2}-Bx.
\end{equation}
This system has two fixed points, $x=0$ and $x=(B/A)^2>0$. A linear stability analysis reveals that the
positive fixed point is linearly unstable, meaning that every initial condition $x_0>(B/A)^2$ will stay
above this value for all times. Further, we know that the solution will grow without bound in this case,
so we can claim the existence of two constants, $t_0<\infty$ and $0<C_0<A$, such that
$Ax^{3/2}(t)-Bx(t)>C_0x^{3/2}(t)$ for every $t>t_0$. This implies that
\begin{equation}
\frac{d}{dt} \left| \left| v \right| \right|_{L^2(\Omega)}^2>
C_0 \left( \left| \left| v \right| \right|_{L^2(\Omega)}^2 \right)^{(3/2)}
\end{equation}
for $t>t_0$,
and for an adecuate initial condition. Solving this equation gives:
\begin{equation}
\left| \left| v(\cdot,t) \right| \right|_{L^2(\Omega)}^2 > \frac{1}
{\sqrt{\left| \left| v(\cdot,t_1) \right| \right|_{L^2(\Omega)}^{-1}-\frac{C_0}{2}t}}
\end{equation}
for $t>t_1>t_0$,
and for an adecuate initial condition. And every adecuate initial condition must fullfill
\begin{eqnarray}
\left| \left| v(\cdot,0) \right| \right|_{L^2(\Omega)}^2 > \frac{4N^2\mu^2}{\epsilon^4}\left| \Omega \right| \nonumber \\
+ \frac{4N\mu}{\epsilon^2} \left| \left| v(\cdot,0) \right| \right|_{L^1(\Omega)}
+ \frac{1}{\left| \Omega \right|} \left| \left| v(\cdot,0) \right| \right|_{L^1(\Omega)}^2,
\end{eqnarray}
like, for instance, $v(x,0)=(x^2 + \delta)^{-1/4} - (L^2 + \delta)^{-1/4}$ and $\delta$ small enough.
Thus we are finally led to conclude that the system does
blow up in finite time. If we recover $v=-\partial_x u$ we see that the first spatial derivative of $u$ becomes singular
in finite time. This means that the solution to
equation Eq.(\ref{eqrosenau}) develops a shock wave in
finite time (or what is the same, a discontinuity in the flow appears),
in contrast to the viscous Burgers' equation and analogously to the inviscid Burgers' equation
$\partial_t v=-v\partial_xv$~\cite{evans}. The inviscid Burguers' equation is a one-dimensional model for the
Euler equations, while the viscous Burguers' equation simulates the Navier-Stokes equations. This suggests that
the regularizating procedure implies a return to a lower order in the Chapman-Enskog expansion.

It was already argued by Rosenau that this kind of regularization of the Chapman-Enskog expansion was only
valid in the linear regime, while nonlinear terms might be present in the full nonlinear hydrodynamics.
These terms are expected to have a deep impact on the dynamics of the fluid, the reason being as follows.
Whether or not the Navier-Stokes equations become singular in finite time is still unkown and it is
actually one of the most important open problems in Mathematics. What one would expect from a physical
point of view is that these possible divergences smooth out if we look closer to the fluid taking into
account higher order terms in a (complete) regularized Chapman-Enskog expansion.
What we have seen in this work is that
the linear regularized theory is able to convert a smooth solution into a singular one, so one would expect
that a regularized Navier-Stokes equation of the form:
\begin{eqnarray}
\nonumber
\partial_t {\bf v} + ({\bf v}\cdot \nabla){\bf v} &=&
-\nabla p +\mu \frac{\nabla^2}{1-m^2 \epsilon^2 \nabla^2}{\bf v} \\
\nabla \cdot {\bf v} &=& 0
\end{eqnarray}
is less regular than the original Navier-Stokes equation. We expect the presence of this nonlinear
terms to regularize enough this equation that one would be able to prove global existence in time of
the solution, and this way give a precise physical meaning to the possible divergences arising in the
original Navier-Stokes equation.

The author gratefully acknowledges illuminating discussions with Antonio C\'{o}rdoba, Diego C\'{o}rdoba, and Francisco Gancedo. This work has been partially supported by the Ministerio de Ciencia y Tecnolog\'{\i}a (Spain) through Project No. BFM2001-0291 and by UNED.

\end{document}